\documentclass[final,5p,times,twocolumn]{elsarticle}

\usepackage{amssymb}
\usepackage{amsmath}
\usepackage{svg}

\newcommand{\g}{$\gamma$}

\journal{Applied Radiation and Isotopes}
\begin{document}

\begin{frontmatter}


\title{First experimental results and optimization study of the portable neutron-gamma imager GN-Vision}

\author[1]{J.~Lerendegui-Marco\footnote{jorge.lerendegui@ific.uv.es}}
\author[1,2]{J. ~Hallam}
\author[1,3]{G.~Cisterna}
\author[4]{A. Sanchis-Molt\'o}
\author[1]{J.~Balibrea-Correa}
\author[1,4]{V.~Babiano-Su\'arez}
\author[1]{D.~Calvo}
\author[1]{I.~Ladarescu}
\author[1]{G.~de~la~Fuente}
\author[1]{B. Gameiro}
\author[1]{P. Torres-S\'anchez}
\author[1]{C.~Domingo-Pardo}

\affiliation[1]{organization={Instituto de Física Corpuscular, CSIC-Universitat de València},country={Spain}}
\affiliation[3]{organization={Universidad de Sevilla},country={Spain}}
\affiliation[2]{organization={University of Surrey},country={United Kingdom}}
\affiliation[4]{organization={Universitat de València},country={Spain}}

\begin{abstract}
GN-Vision is a compact, dual-modality imaging device designed to simultaneously localize the spatial origin of $\gamma$-ray and slow neutron sources, with potential applications in nuclear safety, security, and hadron therapy. The system utilizes two position-sensitive detection planes, combining Compton imaging techniques for $\gamma$-ray visualization with passive collimation for imaging slow and thermal neutrons (energies below 100 eV). This paper presents the first experimental outcomes from the initial GN-Vision prototype, focused on the development of its neutron imaging capabilities. Following this experimental assessment, we explore the device’s performance potential and discuss several Monte Carlo simulation-based optimizations aimed at refining the neutron collimation system. These optimizations seek to improve real-time imaging efficiency and cost-effectiveness, enhancing GN-Vision's applicability for future practical deployments.
\end{abstract}


\end{frontmatter}




\section{Motivation}\label{sec:GNVision}

 In recent years, innovative dual neutron-gamma imaging technologies have garnered significant interest within nuclear security applications, as they enable the detection and spatial localization of a broader spectrum of radioactive sources. These systems are particularly valuable for monitoring spent nuclear fuel in reactors~\cite{Parker:15}, conducting remote inspections in the aftermath of nuclear accidents~\cite{Sato:19,Vetter:18}, and identifying special nuclear materials (SNM) in homeland security applications~\cite{Polack:11,Poitrasson:15,Petrovic:21}. In this context, simultaneous, real-time imaging of $\gamma$-rays and neutrons has gained importance, leading to substantial research efforts focused on developing compact, multimodal imaging devices~\cite{Soundara:12,Whitney:15,Hamrashdi:20,Steinberger:20,Boo:21,Guo:21,Lopez:22}.

In addition to security applications, neutron-gamma imaging systems offer promising solutions for real-time dose monitoring for neutrons and gamma rays~\cite{Schneider:15,Halg:20} and ion-beam range verification~\cite{Durante:19,Pausch:20} in ion beam therapy, an increasingly prevalent cancer treatment technique worldwide. However, the often substantial size of existing devices limits their integration into clinical treatment rooms~\cite{Clarke:16}. Recently, compact dual imaging devices tailored for range verification have been proposed in conceptual studies~\cite{Meric:23,Schellhammer:23}. Neutron and $\gamma$-ray imaging are expected to enable accurate dosimetry in the still experimental Boron Neutron Capture Therapy (BNCT)~\cite{Winkler:15,Hou:22}. 

In response to these challenges, we have recently conceptualized an innovative neutron- and \g-ray-imaging device~\cite{Patent}, hereafter referred to as GN-Vision, that aims to address the most relevant challenges for the applications above. GN-Vision is a compact device capable of simultaneously detecting and spatially localizing \g-rays and slow -- thermal to 100~eV -- neutron sources, both of them with high efficiency~\cite{Lerendegui:22_ANPC,Lerendegui:24}. For the imaging of \g-rays, this device operates as a Compton camera consisting of two position-sensitive detection (PSD) planes. In order to achieve the imaging of slow neutrons, the first position-sensitive detection layer is chosen to have the capability of discriminating $\gamma$-rays and neutrons. A passive neutron collimator attached to this plane allows one to perform neutron imaging with the pinhole camera~\cite{Anger:58,Caballero:18} or coded-aperture mask approaches~\cite{Skinner:84,Cieslak:16}. The latter have been extensively studied in recent years for neutron imaging systems, mostly aimed at fast neutrons~\cite{Griffith:17,Lynde:20}. Among the possible neutron absorbing materials for the collimator~\cite{Stone:19}, lithium polyethylene has been chosen because of the low interference with $\gamma$-rays -- related to its low Z -- and the absence of neutron-induced secondary gamma production.

GN-Vision was first conceptually designed on the basis of Monte Carlo simulations, which demonstrated its simultaneous \g-ray and neutron detection and imaging capabilities~\cite{Lerendegui:22_ANPC,Lerendegui:24}. In this work, we first review the GN-Vision concept and the design of the earliest prototype based on the predecessor i-TED Compton imager, see Sec.~\ref{sec:GNVision}. Following the development of i-TED, the \g-imaging capability in GN-Vision is already at a very high technology readiness level (TRL) following . Thus, the experimental work so far has focused on the neutron imaging module, composed of a position-sensitive CLYC detector -- capable of discriminating \g-rays and neutrons -- and a passive neutron collimator~\cite{Lerendegui:24_arxiv}. In Sec.~\ref{sec:FirstExpResults}, we present the recent progresses in the experimental development leading to the first proof-of-concept (POC) of the neutron imaging capability. Sec.~\ref{sec:Prospects} addresses the future prospects for the final device and outlines several optimization studies aimed at enhancing the performance of the GN-Vision system. Finally, Sec.~\ref{sec:Summary} provides a summary of the results and an outlook for the complete GN-Vision prototype.

\section{The GN-Vision concept and first design}\label{sec:GNVision}

\begin{figure}[!htb]
\centering
\includegraphics[width=0.9\linewidth]{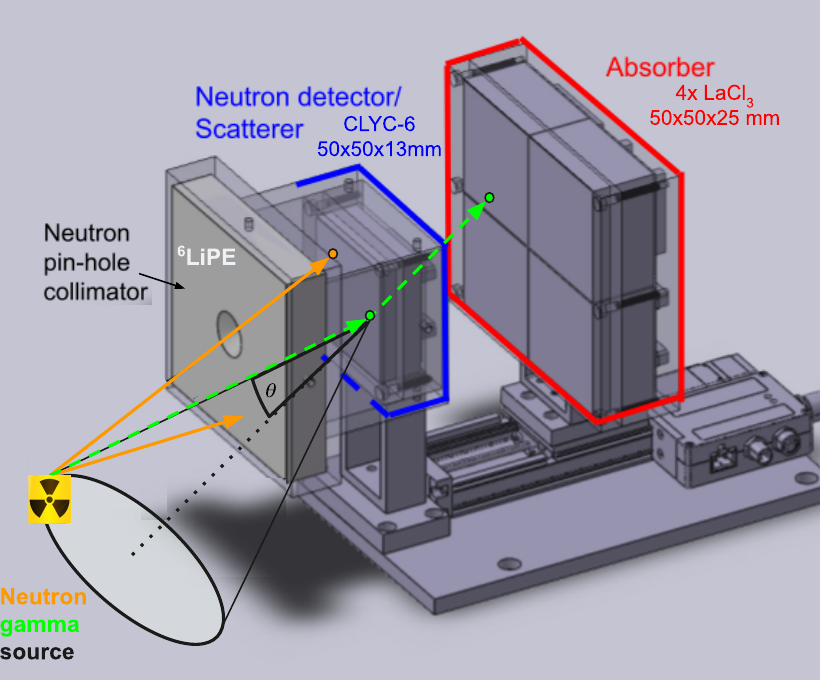}
\caption{Sketch of the earliest GN-Vision prototype showing the working principle of the dual imaging of $\gamma$-rays and neutrons. The main components of the device have been highlighted.}
\label{fig:GNVisionConcept}
\end{figure}

The design of GN-Vision aims to enable exploration of its potential applicability in the different fields and challenges mentioned in the introduction. The system consists of a compact and handheld-portable device capable of measuring and simultaneously imaging \g-rays and slow neutrons, both of them with high efficiency~\cite{Lerendegui:22_ANPC,Lerendegui:24}. The proposed imaging device follows a novel working principle, sketched in Fig.~\ref{fig:GNVisionConcept}:
\begin{itemize}
    \item The Compton imaging technique~\cite{Compton,Domingo16} is exploited to detect and image \g-rays with energies between 100 keV and several MeV using two detection planes, labeled as Scatterer and Absorber in Fig.~\ref{fig:GNVisionConcept}. 
    
   \item The first detection plane is chosen to have the capability of detecting and fully absorbing neutrons of energies $<$1~keV and discriminating them from \g-rays. 
   
   \item  A passive neutron collimation system made out of a material with high absorption power for thermal and epithermal neutrons is attached to the first detection plane and allows to carry out neutron imaging with the same working principle as pinhole~\cite{Anger:58,Caballero:18} or coded-aperture mask~\cite{Skinner:84,Cieslak:16} cameras for $\gamma$-rays.
\end{itemize}

The proposed implementation for the first prototype of GN-Vision, sketched in Fig.~\ref{fig:GNVisionConcept}, represents an evolution of the predecessor i-TED imager~\cite{Domingo16}, which is an array of Compton cameras that has been developed in recent years~\cite{Olleros18,Babiano19,Babiano20,Balibrea:21}. These Compton imagers have been successfully utilized in nuclear astrophysics experiments at CERN n\_TOF~\cite{Babiano:21,Lerendegui:22_NIC, Domingo:22_ND,Lerendegui:23_NPA,Domingo:23} and adapted for medical and nuclear security applications~\cite{Lerendegui:22,Balibrea:22,Lerendegui:24_AppRadIsot,Torres:24,Babiano:24}. i-TED has been modified in such a way that the capability to image $\gamma$-rays is complemented in the same device with the imaging of neutrons. 

For the imaging of $\gamma$-rays, GN-Vision functions in the same manner as the i-TED concept, as a Compton camera consisting of two PSDs~\cite{Domingo16}. The fourfold increase in the size of the absorber relative to the scatter enables a superior performance in terms of efficiency, especially for the detection of low-energy (100-500~keV) gamma rays. The prototype incorporates an embedded micro-positioning drive, which enables variable adjustment of the distance between the two detection planes. In order to achieve the simultaneous imaging of neutrons, the LaCl$_{3}$ crystal of the first detection layer of i-TED is replaced by a Cs$_{2}$LiYCl$_{6}$:Ce scintillation crystal enriched with $^{6}$Li at 95\% (CLYC-6), which is capable of discriminating $\gamma$-rays, fast and thermal neutrons by Pulse Shape Discrimination (PSD)~\cite{Giaz:16}. The use of a CLYC-6 scintillator as a scatterer detector ensures also sufficient energy resolution for Compton imaging in GN-Vision. As shown in Fig.~\ref{fig:GNVisionConcept}, in its first design, GN-Vision comprises a neutron absorbing pinhole collimator made of $^{6}$Li-enriched Lithium Polyethylene ($^6$LiPE), which is coupled to the first detection plane to allow for the reconstruction of 2D neutron-images. On the other hand, the $^6$LiPE material does not interfere with $\gamma$-rays above $\sim$100keV, thereby enabling the simultaneous neutron-gamma vision within the same device. 

\section{First experimental results} \label{sec:FirstExpResults}

\subsection{Experimental development and characterization} \label{sec:ExpDevelopment}

GN-Vision was first conceptually designed on the basis of Monte Carlo (MC) simulations which demonstrated its simultaneous \g-ray and neutron detection and imaging capabilities~\cite{Lerendegui:24}. In this work, we optimized the key parameters of the $^6$LiPE neutron pinhole collimator and studied the neutron imaging performance for the range of energies of interest (thermal - 100 eV). Moreover, this work proved the ability of the GN-Vision concept for the Compton imaging of \g-ray energies as low as 500~keV despite the presence of a neutron collimator. Similar MC-based studies are presented in this work aimed at further optimizing the performance of the complete GN-Vision prototype, see Sec.~\ref{sec:Prospects}.

\begin{figure*}[!t]
\centering
\includegraphics[width=0.7\linewidth]{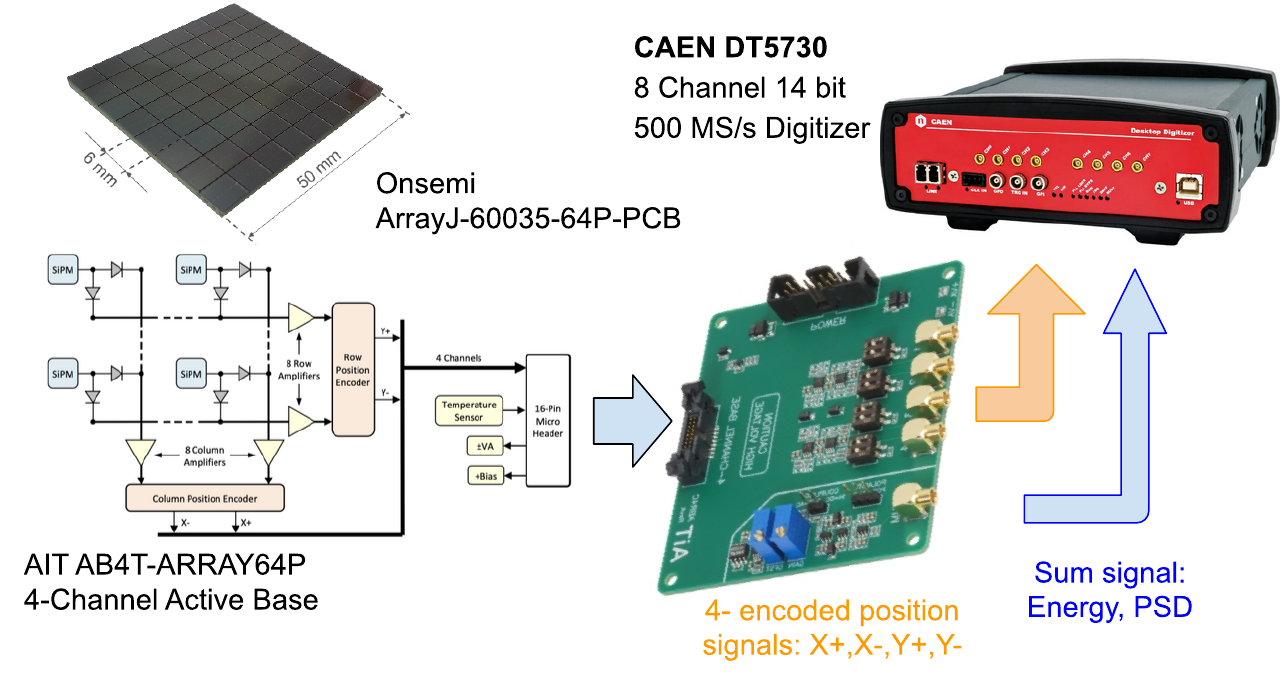}
\caption{Components and electronic chain from the SiPM to the signal processing with the CAEN DACQ used in the development and characterization of the position-sensitive CLYC detector of GN-Vision.}
\label{fig:SketchElectronics}
\end{figure*}

Empowered by the satisfactory results of the conceptual study, the experimental development to date has been focused on the central component of GN-Vision, its first position sensitive detection layer featuring particle discrimination capability. A monolithic CLYC-6 crystal with dimensions of 50$\times$50$\times$13~mm was chosen for this purpose~\cite{Lerendegui:24_arxiv}. With the aim of deploying a position sensitive detector (PSD), the crystal was coupled to the Onsemi SiPM ARRAYJ-60035-64P, the same model used in the i-TED detector~\cite{Babiano19,Balibrea:21}. This photo-sensor features 8$\times$8 pixels over a surface of 50$\times$50 mm$^{2}$. The readout of the CLYC-SiPM module, sketched in Fig.~\ref{fig:SketchElectronics}, was carried out with the AIT 4-Channel Active Base (AB4T-ARRAY64P)~\cite{AIT} frontend. This readout electronics enables the implementation of a multiplexing method (i.e. Anger Logic)~\cite{Anger:58}, to extract four encoded position signals out of the 64 signals of the individual SiPMs. The output signals, comprising four weighted position signals and a summed signal for spectroscopy and pulse shape discrimination (PSD), were fed into a CAEN DT5730S digitizer, which acquires analogue signals via Flash ADCs with 14-bit resolution and 500 MS/s sampling rate and performs online digital pulse processing (DPP).

 As part of the development of the CLYC-SiPM detector of GN-Vision, we characterized the energy resolution and performance in terms of PSD of neutrons and $\gamma$-rays. An energy resolution of 6.2\% (662 keV) was found with a PMT readout, that degraded to 8.9\% when the crystal was instead readout via the SiPM array and the sum signal of the 64 pixels was used. In terms of PSD the CLYC-6 provides a great separation between thermal neutrons and $\gamma$-rays, characterized by FOM values of 3.8 and 2.9, respectively, with a PMT and SiPM readout. The characterization of the spatial response of the CLYC-SiPM assembly has delivered a sub-pixel position resolution of 5~mm for the SiPM array of 6~mm size pixels. Moreover, a linear response has been found across the central 30~mm region of the crystal. For more details on the experimental development and characterization of the CLYC-SiPM detector, the reader is referred to Ref.~\cite{Lerendegui:24_arxiv}.
 
 The complete experimental characterization of the position-sensitive CLYC detector presented herein, together with previous characterization works on the LaCl$_3$ detectors of the absorber plane~\cite{Babiano19,Balibrea:21}, have allowed us to study in a realistic manner the expected performance of the GN-Vision device, which is discussed in Sec.~\ref{sec:Prospects}. The following steps in the development of the proposed device will cope with the integration of the dual imaging technique. The combination of both imaging modalities will require the integration in the same readout system of the position-sensitive CLYC detector and the position-sensitive absorber plane based on LaCl$_3$ crystals. More details on the ongoing research are given in the outlook of this paper in Sec.~\ref{sec:Summary}.


\subsection{Neutron imaging proof-of-concept} \label{sec:NeutronImaging}


\begin{figure}[!b]
\centering
\includegraphics[width=0.9\linewidth]{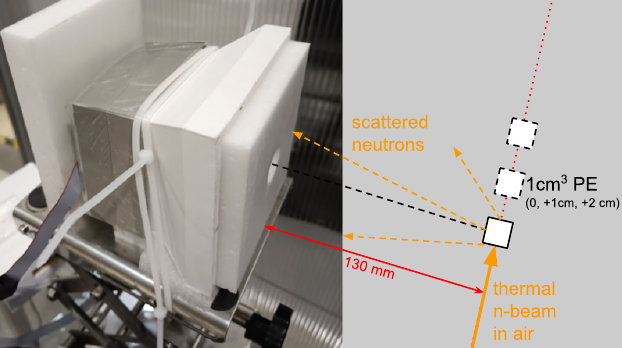}
\caption{Experimental setup for the proof-of-concept experiments of neutron imaging. The neutron imaging module was placed in front of PE cubes that were irradiated in different positions along the beam axis.}
\label{fig:Imagingsetup}
\end{figure}

The development and characterization of the position-sensitive neutron-gamma discriminating CLYC detector has laid the foundation for the first proof-of-concept experiment of the neutron imaging capability of GN-Vision. For such an experiment, a neutron imaging module was assembled with the CLYC-SiPM detector and a $^6$LiPE neutron pinhole collimator following the first GN-Vision design (Fig.~\ref{fig:GNVisionConcept}). The first POC experiments were carried out at ILL-Grenoble using the scattering of a thermal neutron beam in small plastic targets~\cite{Lerendegui:24_arxiv}.

\begin{figure}[!t]
\centering
\includegraphics[width=0.75\linewidth]{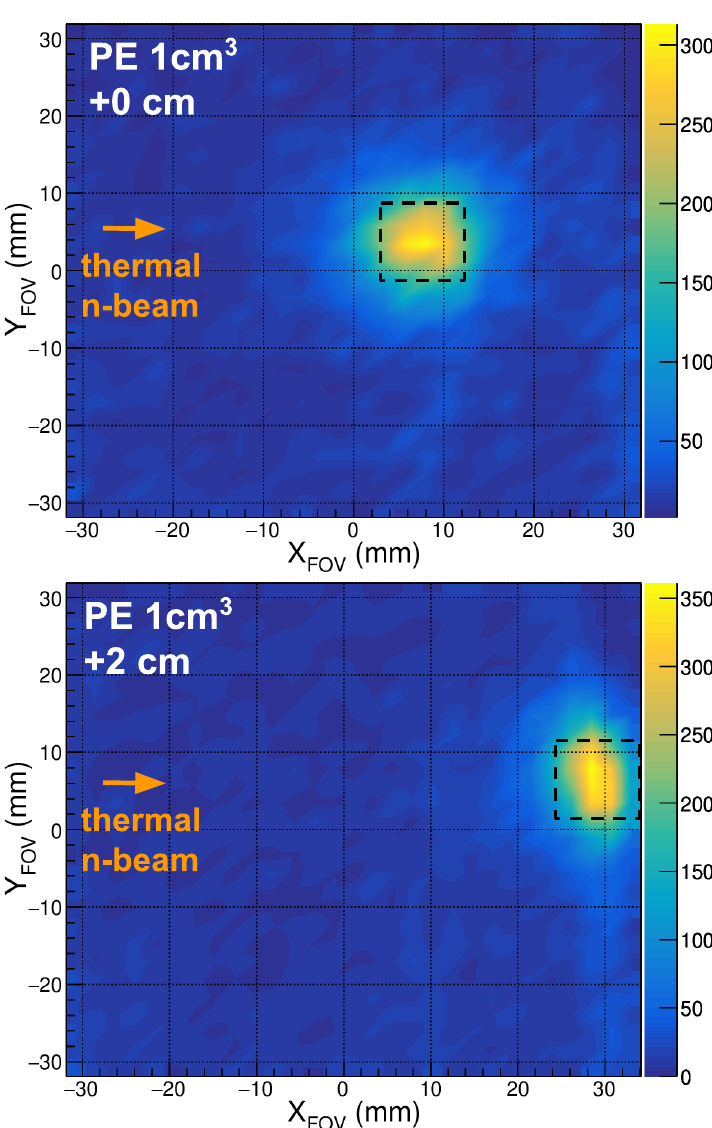}
\caption{First experimental neutron images reconstructed with GN-Vision using thermal neutrons scattered in a polyethylene cube of 1~cm$^{3}$ placed in two different positions. The size of the PE cube is indicated with the dashed box.}
\label{fig:NeutImages}
\end{figure}

The neutron imaging module was installed in the FIPPS experimental hall at the Institut Laue-Langevin (ILL). A neutron beam of 1.5~cm in diameter and a flux of 5$\times$10$^{7}$n/cm$^{2}$/s~\cite{Michelagnoli:18} was used to irradiate a single Polyethylene (PE) cube of 1~cm$^{3}$ size located at a distance of 130~mm from the collimator, see Fig.~\ref{fig:Imagingsetup}. The critical parameters of the neutron imaging device, sketched in Fig.~\ref{fig:MCSetup}, are the thickness (T), diameter (D) and focal distance (F) of the pinhole. In the prototype used for this POC experiment T = 20 mm, D = 2.5 mm and F = 40~mm. The reader is referred to the previous MC-based study~\cite{Lerendegui:24} for the details on the impact of these parameters on the imaging performance and to Ref.~\cite{Lerendegui:24_arxiv} for more details on the experimental campaign. 

The first neutron images reconstructed with GN-Vision are shown in Fig.~\ref{fig:NeutImages}. These images were generated from the 2D-coordinates of the selected slow neutron hits simply by applying inversions in both the $x$ and $y$ planes and a scaling factor $S = d / F$, where $d$ is the distance from the collimator pinhole to the neutron source plane (i.e. 130 mm, see Fig.~\ref{fig:Imagingsetup}), and $F$ is the focal distance of the pinhole. The top and bottom panels of Fig.~\ref{fig:NeutImages}  show two examples of the obtained neutron images, corresponding to the PE cube roughly centered with respect to the detector and shifted 2~cm upstream. A clear pattern is observed in the images which follows the expected shift  of the PE cube (indicated with the black dashed line) and has the expected dimensions (11~mm FWHM). A remarkable contrast -- peak-to-background of $\sim$15 -- has been obtained. Besides the imaging examples displayed in Fig.~\ref{fig:NeutImages}, the experimental campaign at ILL served to demonstrate the remarkable angular resolution below 5$^{\circ}$ achieved with the earliest prototype of the neutron imaging block of GN-Vision~\cite{Lerendegui:24_arxiv}. 

The successful results of the neutron imaging proof-of-concept experiment reviewed in this section represent a major milestone in the experimental development of the GN-Vision device. At present, the main limitations that we have identified in the first GN-Vision prototype are mainly related to the neutron-detection efficiency and the commercial availability and cost of $^{6}$Li-enriched material for the neutron collimation system. To tackle these challenges, we have recently investigated the possibility of enhancing the efficiency in at least one order of magnitude by using a coded-aperture neutron mask. In addition, to make GN-Vision a more cost-effective solution, the impact of replacing the 95\% enriched $^{6}$LiPE by natural LiPE has been analyzed for various applications. These two topics are presented in Sec.~\ref{sec:Prospects}.



\section{Prospects of the final device} \label{sec:Prospects}
\subsection{Performance and impact of the resolution}\label{sec:MCPerformance}

The performance of GN-Vision is determined by several features, such as energy- and position-resolution.
In particular, in this section we study the effect of the spatial and energy resolution of the CLYC-based position-sensitive layer. For the absorber detection plane, which is based on LaCl$_{3}$ crystals, the impact of the energy and spatial resolutions on the image resolution is well characterized from previous works of the predecessor i-TED detector~\cite{Babiano19,Balibrea:21,Gameiro:23}. 

\begin{figure}[!b]
\centering
\includegraphics[width=1\linewidth]{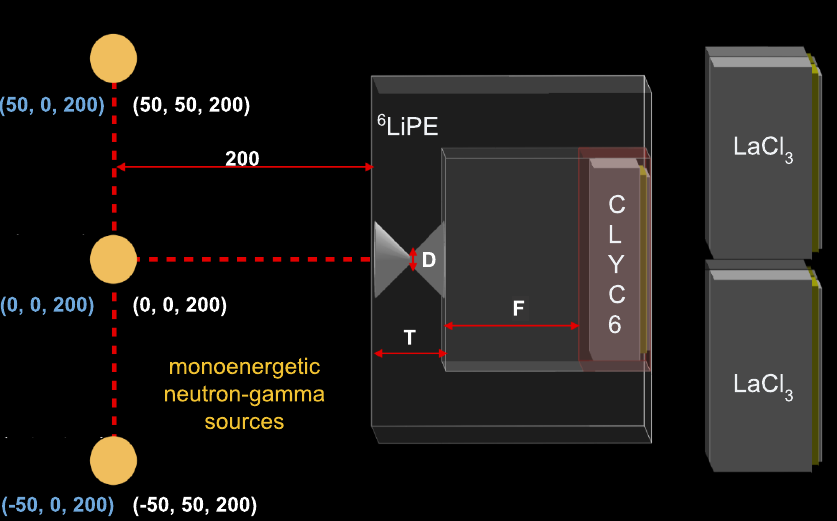}
\caption{Geometry model of GN-Vision as implemented in Geant4. The main components of the device, relevant parameters of the pinhole collimator and coordinates of the primary simulated sources are indicated. See text for details.}
\label{fig:MCSetup}
\end{figure}

\begin{figure*}[!t]
\centering
\includegraphics[width=0.65\linewidth]{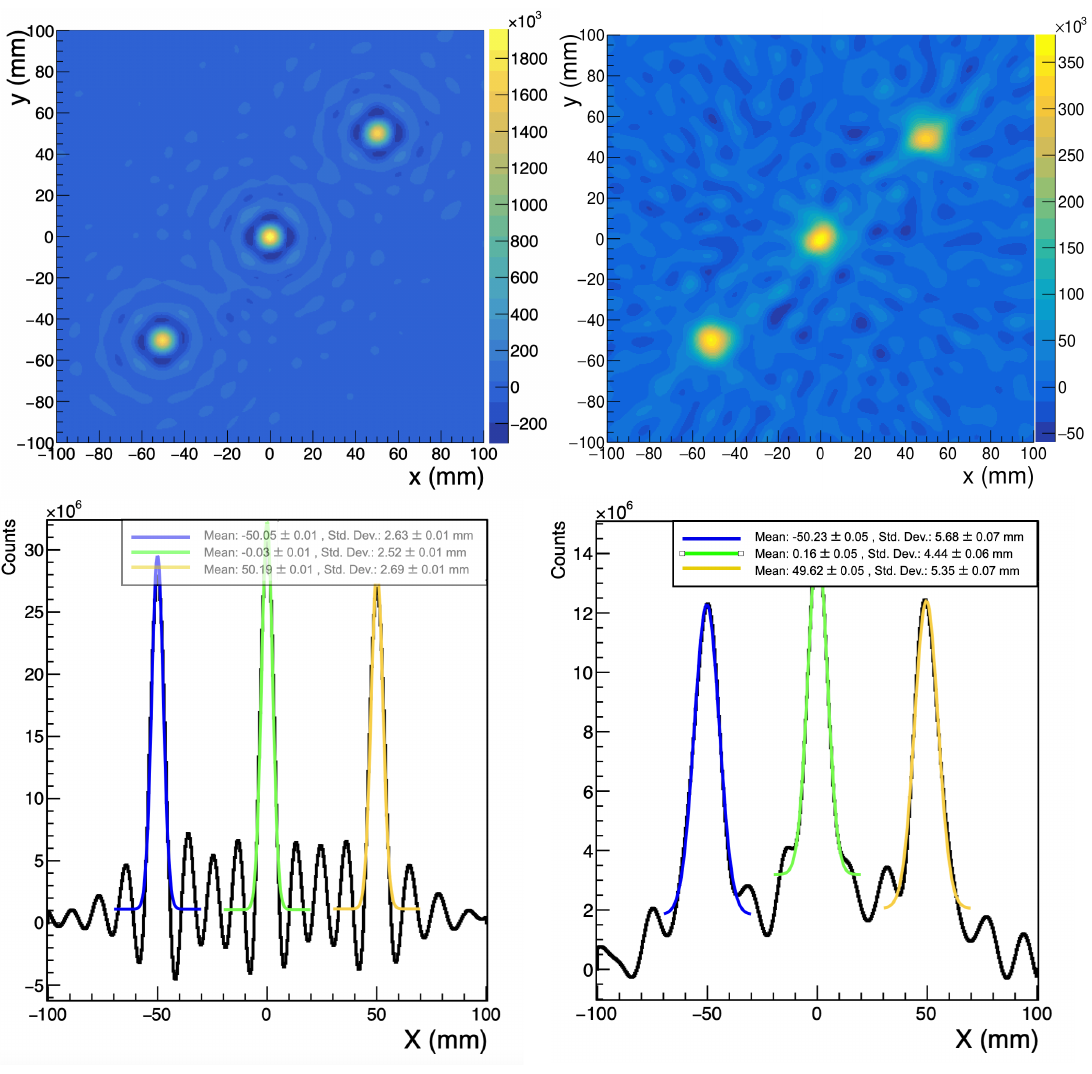}

\caption{Compton images resulting from the simulations of GN-Vision response to 3 point-like sources of 511~keV $\gamma$-rays (top) and projections onto the X-axis (bottom). The ideal results from simulations (left) are compared to the realistic performance including experimental effects (right).}
\label{fig:ComptonImpactExpResol}
\end{figure*}

To evaluate the expected performance of the utter GN-Vision prototype, the responses of the system to both \g-rays and neutrons were generated by means of Monte-Carlo simulations carried out with the \textsc{Geant4} toolkit~\cite{Geant4_2} using the officially released QGSP\_INCLXX\_HP Physics List~\cite{Geant4PL}. For an accurate simulation of the neutron interactions, neutron-induced reactions below 20 MeV are simulated within this \textsc{Geant4} PL by means of the G4NeutronHP package~\cite{Mendoza:14}, using the G4NDL-4.6 data library (based on the JEFF-3.3~\cite{Plompen:20} evaluated data file). The geometry model implemented in \textsc{Geant4} is shown in Fig.~\ref{fig:MCSetup}. In the simulations,
the neutron-collimator design parameters (D, T and F in Fig.~\ref{fig:MCSetup}) were set to the same values of the prototype used for the neutron imaging POC experiment discussed in Sec.~\ref{sec:NeutronImaging}. Three isotropic point-like sources of either \g-rays or neutrons were placed at a distance of 200~mm in three different positions along the diagonal of its FOV. In Fig.~\ref{fig:MCSetup}, the exact coordinates of the sources are indicated in white. Neutrons with an energy of 0.025 eV, resembling the conditions of the first POC experiment at ILL, and \g-rays of 511 keV, such as the ones used for the experimental characterization of the spatial resolution of both the CLYC and LaCl$_3$ detectors~\cite{Balibrea:21,Lerendegui:24_arxiv}, were emitted isotropically from these sources. 

The output of the MC simulation features the same format as the experimental data, including for each simulated event the deposited energy, interaction position and time of all the neutron and \g-ray hits in the two position sensitive layers of GN-Vision. To mimic the experimental discrimination of \g-rays and neutrons via pulse shape discrimination in the CLYC crystal, see Ref.~\cite{Lerendegui:24_arxiv}, a flag was included to identify energy depositions via $^{6}$Li(n,$\alpha$)$^{3}$H reactions from those carried out by electrons associated to \g-ray events.

The $\gamma$-ray images were reconstructed via the Compton technique, in which only $\gamma$-ray hits in time coincidence between the Scatter (S) and Absorber (A) position-sensitive detectors (see Fig.~\ref{fig:GNVisionConcept}) are considered from the output of the MC simulations. More details on the Compton technique and the implementation of the imaging algorithms can be found elsewhere~\cite{Babiano20,Babiano:21,Lerendegui:22,Balibrea:22}. Fig.~\ref{fig:ComptonImpactExpResol} shows the \g-ray images obtained with GN-Vision for three point-like sources of 511~keV \g-rays placed in a plane at 200~mm of the device and in three positions along the diagonal. The analytical algorithm of Tomitani et al.~\cite{Tomitani:02}, implemented already for i-TED in~\cite{Lerendegui:22,Balibrea:22}, has been used to reconstruct these images. In the case of slow neutrons, the simple pinhole imaging algorithm~\cite{Lerendegui:24} used for the experimental validation of the neutron imaging in Sec.~\ref{sec:NeutronImaging} was applied.

\begin{figure*}[!tb]
\centering
\includegraphics[width=0.65\linewidth]{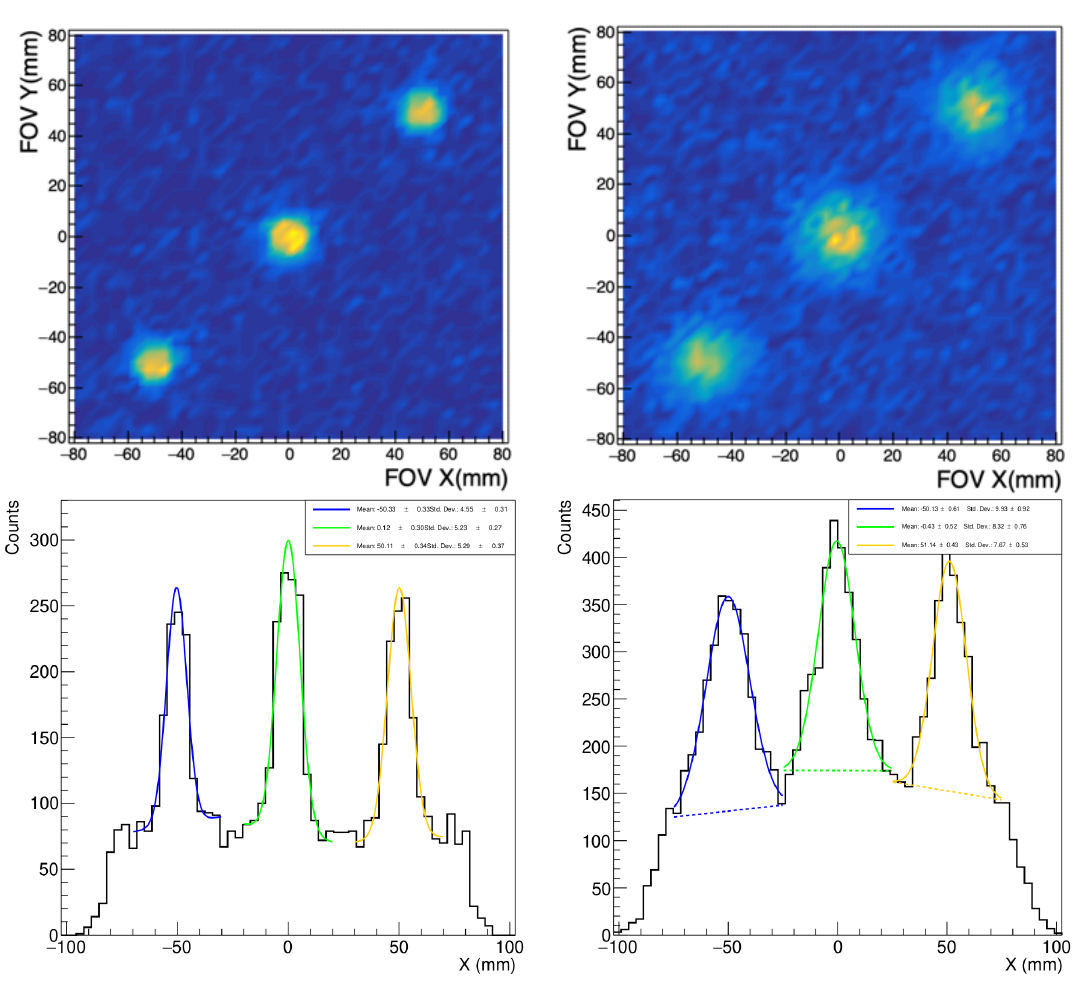}
\caption{Neutron images resulting from the simulations of GN-Vision response to 3 point-like sources of 0.025~eV neutrons (top) and projections onto the X-axis
(bottom). The ideal results from simulations (left) are compared to the realistic performance including experimental effects (right).}
\label{fig:NeutronImpactExpResol}
\end{figure*}

The resulting Compton image for the three $\gamma$-ray sources reconstructed with an ideal implementation of the GN-Vision device (i.e. with no experimental effects) is shown in the top left panel of Fig.~\ref{fig:ComptonImpactExpResol}. The panel below corresponds to the projections of the 2D image onto the X axis. Left panels of Fig.~\ref{fig:NeutronImpactExpResol} show the corresponding 2D neutron image and its 1D projection, reconstructed for the 3 point-like neutron sources. Thus, the 1D projections shown for both $\gamma$-ray and neutrons serve to quantify the ideal detector performance, in terms of linearity and resolution, for the chosen GN-Vision configuration.

The impact of the experimental spatial- and energy-resolutions in the neutron-gamma imaging was investigated by adding these experimental effects to the output of the simulations before the application of the image-reconstruction algorithms.  As for the CLYC crystal -- acting as neutron detector and Compton scatter plane -- the coordinates in the transverse plane of both neutron and $\gamma$-ray hits were convoluted with a Gaussian distribution of 5~mm (FHWM) in agreement with our experimental findings~\cite{Lerendegui:24_arxiv}. In the case of the z coordinate, the interaction position has been fixed at the middle of the crystal thickness since the actual multiplexing approach used so far to read out the CLYC-SiPM detector is not sensitive to the depth-of-interaction (doi). Regarding the LaCl$_3$ crystals -- of use in the Compton absorber plane -- the spatial resolution in the transverse coordinates and doi was, respectively, 1.5 mm and 4~mm~\cite{Balibrea:22}. The compression effects, discussed in detail in Refs.~\cite{Balibrea:22,Lerendegui:24_arxiv}, were also considered resulting in a shift of the reconstructed position in the peripheral region towards the central part of the crystal. An energy resolution of 6\% at 662 keV was assumed with an energy dependence of \(\sqrt{1/E}\). This energy resolution is not far from that reported for both the CLYC and LaCl$_3$ crystals~\cite{Lerendegui:24_arxiv,Gameiro:23}. Last, a realistic threshold of 100~keV in deposited energy per crystal has been assumed. 

After adding the aforesaid experimental effects, the resulting Compton images are shown in the right panels of Fig.~\ref{fig:ComptonImpactExpResol}. From the comparison of the images in Fig.~\ref{fig:ComptonImpactExpResol}, we conclude that the image resolution worsens by a factor of almost two due to the experimental effects. In contrast, the centroids of the sources do not change much with these spatial effects, maintaining the values of the position of the source within the error. A side effect of the lower resolution is the increase of the background and consequently the loss of contrast -- defined as peak-to-background ratio -- when compared to the ideal device with no experimental spatial effects.  In a similar fashion, the right panels of Fig.~\ref{fig:NeutronImpactExpResol} show the neutron image that is expected to be reconstructed taking into account the spatial resolution and non-linear response near the crystal edges of the CLYC-SiPM detector. As in the gamma image, we observe a two-fold broadening of the point source resolution as a result of the experimental effects. The image resolution of about 8~mm ($\sigma$)-- for sources at 210~mm from the center of the collimator -- corresponds to an angular resolution of $\approx$5$^{\circ}$, which is in very good agreement with the one experimentally demonstrated in the first proof-of-concept experiments also for thermal neutrons~\cite{Lerendegui:24_arxiv}.


\subsection{Collimator material: a cost-effective solution}\label{sec:MCCollimatorMat}

 The design of the first GN-Vision prototype comprises a simple pinhole collimator for the imaging of slow neutrons. In the design phase of GN-Vision, we chose, among other low-Z neutron-absorbing materials, highly (95\%) $^{6}$Li-enriched polyethylene ($^6$LiPE)~\cite{Lerendegui:24}. The high enrichment in $^{6}$Li, compared to its natural abundance of 7.4\%, was chosen to boost the performance of the collimator due to the large neutron absorption of this isotope and has been proven to provide very nice resolution and contrast results, as presented in Sec.~\ref{sec:NeutronImaging}. However, the need to enrich the raw material to produce the LiPE neutron collimator poses some challenges related to the cost-effectiveness and scalability of the proposed device. Moreover, the initial MC-based study indicated that for the imaging of thermal neutrons a very thin layer of $^6$LiPE would be sufficient to obtain a satisfactory contrast~\cite{Lerendegui:24}, opening the door to a lower enrichment -- and subsequently reduced cost -- in future prototypes. For all of the above, we have recently explored, on the basis of MC simulations, the neutron imaging performance of GN-Vision with a pinhole collimator produced out of natural-Li polyethylene ($^{nat}$LiPE) and compared it to the baseline results with enriched material.

\begin{figure*}[!htb]
\centering
\includegraphics[width=0.8\linewidth]{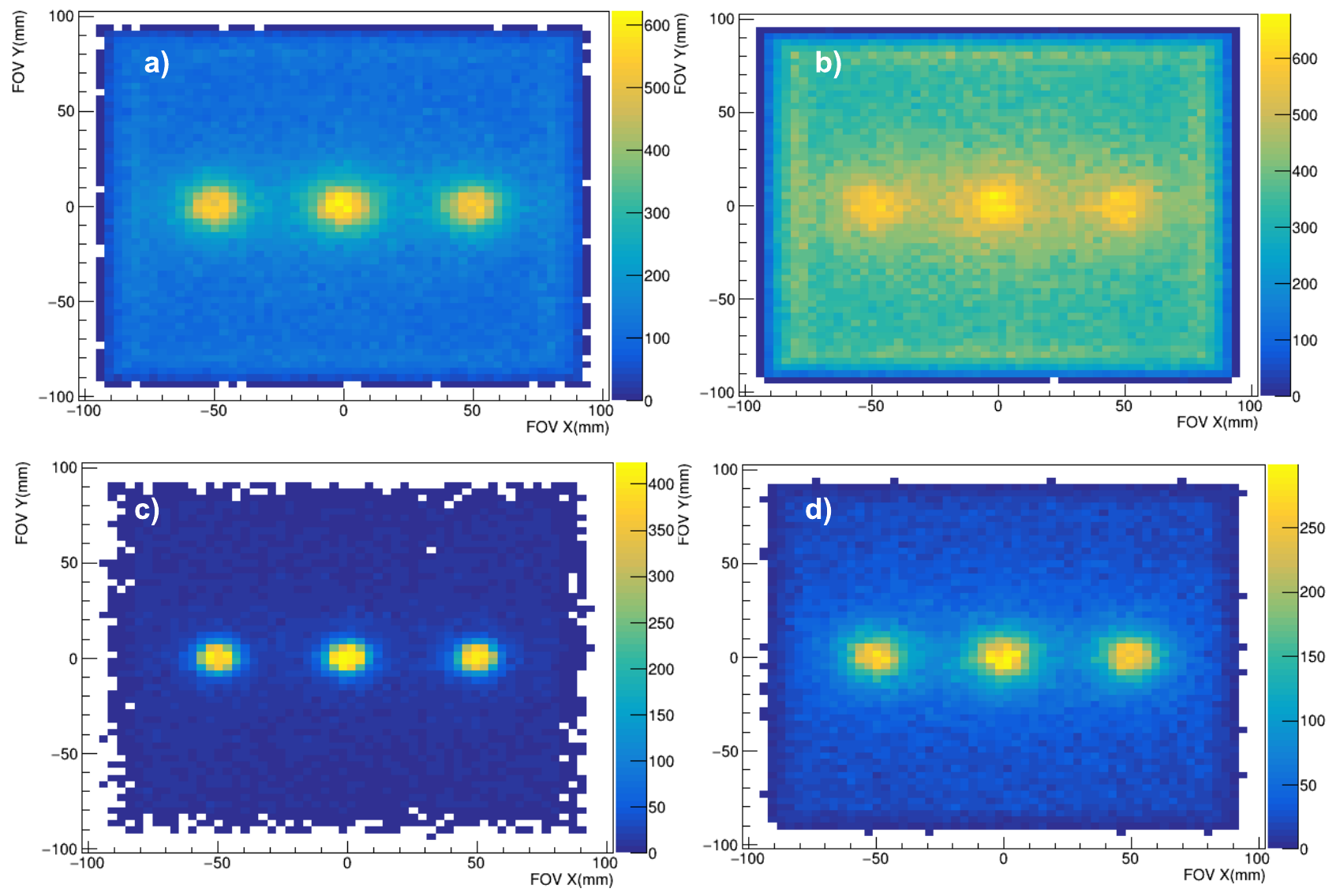}
\caption{Reconstructed neutron images for three isotropic point-like sources neutrons located at 20 cm from GN-Vision with a collimator of T=20~mm,D=2.5~mm, F=40~mm. a) and c) correspond to thermal neutrons and b) and d) to a neutron energy of 1~eV. a) and b) show the results obtained for $^{nat}$LiPE while c) and d) for $^{6}$LiPE.}
\label{fig:NeutronCollimatorMaterial}
\end{figure*}

Simulations were performed using the geometry model of GN-Vision implemented in \textsc{Geant4} sketched in Fig.~\ref{fig:MCSetup}.  In these simulations, the diameter and focal distance of the pinhole collimator (D, T in Fig.~\ref{fig:MCSetup}) were set again to the same values of the POC set-up (Sec.~\ref{sec:NeutronImaging}) and in the MC-based performance study (Sec.~\ref{sec:MCPerformance}). Collimator thicknesses ranging from 10 to 30~mm were simulated for both $^{nat}$LiPE and $^{6}$LiPE to evaluate their performance.  Three isotropic point-like sources of neutrons were located in a plane at 200~mm of the device and in three positions along the horizontal axis, indicated with the blue coordinates in Fig.~\ref{fig:MCSetup}. The simulated neutrons were emitted with two different neutron energies, thermal energies (25 meV) and 1~eV to evaluate the effectiveness of both collimation materials at different neutron energies. As in Ref.~\cite{Lerendegui:24}, the latter energy has been chosen as a representative energy value of epithermal neutrons within the operational range of GN-Vision. The methodology to select neutron events in the simulation and build the corresponding neutron image was the same as the one explained in the previous section.

\begin{table}[htb!]
\centering
\caption{Contrast of the neutron image (PBR) as a function of the neutron energy, neutron collimator material and thickness (T).}
\label{tab:Contrast}       
\begin{tabular}{ccccc}
\hline
Material & En (eV)& T = 10 mm & T = 20 mm   & T = 30 mm\\
\hline
$^6$LiPE&  0.025 &  14(1)   &   15(1)     & 18(2)   \\
$^6$LiPE&  1     &   1.97(6)     &   4.2(2)    &   6.0(4)      \\
$^{nat}$LiPE&  0.025 &  1.65(3)  &  2.76(8)   &  3.8(1)  \\
$^{nat}$LiPE&  1    &   1.22(2)   &  1.42(3)    &  1.65(4)   \\
\hline
\end{tabular}
\end{table}

Fig.~\ref{fig:NeutronCollimatorMaterial} shows some examples of the neutron images reconstructed from the simulated data, which illustrate the impact in the imaging performance of the different neutron energies and collimator materials for a collimator thickness T = 20~mm. Aiming at a quantitative and comprehensive overview of the performance of GN-Vision for all the scenarios under study, we have quantified the resolution and contrast, defined as the peak-to-background ratio (PBR), from the analysis of the image projections onto the x-axis. Table~\ref{tab:Contrast} presents the resulting contrast for the neutron images reconstructed for thermal and 1 eV neutron sources and the different collimator thicknesses (T) under study. The results indicate that a thin $^{6}$LiPE of only 10~mm would be sufficient to achieve an excellent contrast PBR=15 for thermal neutrons while at least 20~mm of $^{6}$LiPE would be required to reconstruct an image with reasonable contrast, for instance PBR $>$2, for a neutron energy of 1~eV. When comparing the performance of the natural and enriched materials, we conclude that collimators made out of 20 or 30~mm of $^{nat}$LiPE would still provide satisfactory contrasts for applications where thermal neutrons dominate the spectrum. In contrast, to image epithermal neutrons of 1~eV  energy, $^{6}$LiPE becomes significantly more convenient, at least to achieve PBR $>$2, regardless of the collimator thickness.  

\begin{table}[htb!]
\centering
\caption{Resolution of the neutron image (FWHM in mm) as a function of the neutron energy, neutron collimator material and thickness (T).}
\label{tab:Resol}       
\begin{tabular}{ccccc}
\hline
Material & En (eV)& T = 10 mm & T = 20 mm   & T = 30 mm\\
\hline
$^6$LiPE&  0.025 &   12.9(2)  & 12.8(4)      &  12.1(2)  \\
$^6$LiPE&  1     & 22(2)     &  19.7(5)    &  20.2(5)       \\
$^{nat}$LiPE&  0.025 &  20.9(6)  & 19.5(4)     &  19.7(3)  \\
$^{nat}$LiPE&  1     &  22(2)     &  23(1)    &  23(1)    \\
\hline
\end{tabular}
\end{table}

The impact of the collimator material in the attainable image resolution is presented in Table~\ref{tab:Resol}. The results indicate that $^6$LiPE provides better resolution for both thermal (25.3~meV) and epithermal (1~eV) neutrons. The difference is more sizable for thermal neutrons, for which the use of $^6$LiPE prevents that the pinhole can be partially transversed by neutrons, an effect that leads to an apparent increase of the effective pinhole diameter (D) and worsens the spatial resolution by $\sim$50\%. For epithermal neutrons, exemplified with $E_n$=1~eV in this work, the pinhole is not totally opaque anymore regardless of the material and for this reason, the spatial resolution varies only by $\sim$10\% between the LiPE with natural and $^{6}$Li-enriched compositions, regardless of the thickness T.


\subsection{Coded-aperture mask: enhancing the efficiency}\label{sec:CodedMask}

As discussed in the conceptual design of GN-Vision~\cite{Lerendegui:24}, one of the main limitations of the first prototype based on a simple pinhole collimator is its limited efficiency for the imaging of neutrons, which is of the order of 10$^{-5}$. A higher detection efficiency to detect and image neutrons would be desirable to maximize the performance of GN-Vision in various practical scenarios, such as nuclear inspections or waste characterization, where the neutron fluence reaching the detector is small. Also, for medical applications requiring real-time imaging during treatment, high detection efficiency is particularly important.

\begin{figure}[h]
\centering
\includegraphics[width=0.7\linewidth]{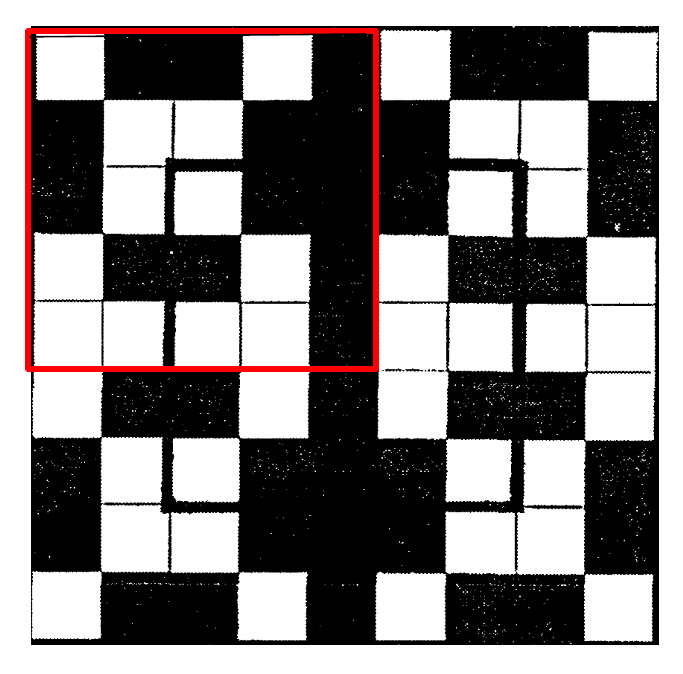}
\caption{The Rank 5 coded mask pattern highlighted in red each the 5$\times$5 aperture construction of Eq.~\ref{eq:Mura5}~\cite{Gottesman89}.}
\label{fig:Rank5}
\end{figure}

A more sophisticated approach to passive collimation imaging involves the use of a coded pattern with a significantly larger open fraction than that of a pinhole collimator, which is considered an optimal configuration. In this section we use MC simulations in order to investigate a potential collimator design, which could be realistically manufactured and physically implemented within a more evolved prototype version of the GN-Vision concept. The coded aperture selected for this study is the Modified Uniformly Redundant Arrays (MURAs)~\cite{Gottesman89}. Owing to their square shape, MURAs offer a diverse range of possible sizes, with an open fraction that tends towards 1/2 for large values of the rank. The results presented hereafter correspond to an MURA aperture pattern designated as Rank 5, whose construction is given by
\begin{align}
b=\begin{vmatrix}
0 & 0 & 0 & 0 & 0 \\
1 & 1 & 0 & 0 & 1 \\
1 & 0 & 1 & 1 & 0 \\
1 & 0 & 1 & 1 & 0 \\
1 & 1 & 0 & 0 & 1
\end{vmatrix},
\label{eq:Mura5}
\end{align}
where 0 and 1 represent solid blocks and holes, respectively. A mosaicked representation of the resulting $2\times2$ symmetrical pattern, which provides an open fraction $\rho = 0.48$, is depicted in Figure \ref{fig:Rank5}. This pattern was selected due to its simpler design when compared with higher MURA ranks. This aspect will  facilitate its machining in LiPE material for the new neutron collimator version of GN-Vision. A more general study for different MURA ranks can be found in Ref.~\cite{JamesThesis}.

The geometry of the coded-aperture mask of Figure~\ref{fig:Rank5} was implemented in the Geant4 application described in previous sections, where it replaced the pinhole collimator shown in Fig.~\ref{fig:MCSetup}. A thin (T = 5~mm) PE$^{6}$Li coded aperture mask with the design of Fig.~\ref{fig:Rank5} was placed in front of the first detection plane of GN-Vision, at a (focal) distance F = 14 mm. This collimator thickness was chosen taking into account that only thermal neutrons would be simulated. As explained in a previous section, only neutron events are selected from the output of the MC simulations.

From the simulated responses to neutron sources located in front of the detector, deconvolution algorithms were employed for unfolding the spatial patterns of the neutron sources. The selected approach involves comparing a given detector response with a database of simulated responses corresponding to a source in a specific location. Two deconvolution algorithms were implemented for image reconstruction in this study, each with distinct characteristics.
\begin{itemize}
    \item The Agostini algorithm aims at estimating the true underlying distribution from the observed data convolved with a response matrix $R$. At each iteration $k$, the algorithm updates the probability estimates $\mathbf{P}^{(k)}$ associated with each bin in the true underlying distribution:
\[
    P_i^{(k)} = \frac{R_{ij} P_j^{(k-1)}}{\sum_j R_{ij} P_j^{(k-1)}}.
\]
The iterative process continues until a convergence criterion is met, and is done so by monitoring the chi-square value:
\[
\chi^2 = \sum_i \left( n_i - n_{0i} \right)^2,
\]
where $n_i$ is the observed data and $n_{0i}$ is the unfolded data.
 \item The Maximum Entropy algorithm aims to reconstruct an estimate of the true underlying distribution from the convolved data, considering the response matrix $R$. The algorithm iteratively refines the probability estimates associated with each bin in the true underlying distribution, seeking to maximize entropy.

At each iteration $k$, the probability estimates ($P^{(k)}$) are updated based on the following equation:
\[
P_i^{(k)} = P_i^{(k-1)} \times \exp\left(\frac{2}{\lambda} \sum_j \frac{R_{ij}(d_i - \sum_k R_{kj} P_k^{(k-1)})}{\sigma_j^2}\right).
\]
Here, $P_i^{(k)}$ represents the probability estimate for bin $i$, $\lambda$ controls the trade-off between fitting the data and maximizing entropy, $d_i$ is the observed data in bin $i$, $R_{ij}$ is the response matrix element, and $\sigma_j$ is the error associated with the observed data in bin $j$. The algorithm continues iterating until a convergence criterion is met or a maximum number of iterations is reached. The convergence is monitored using the $\chi^2$ value:
\[
\chi^2 = \sum_i (P_i - P_i^{(0)})^2
\]
where $P_i$ is the current probability estimate, and $P_i^{(0)}$ is the initial estimate.

\end{itemize}

\begin{figure}[!b]
\centering
\includegraphics[width=0.8\linewidth]{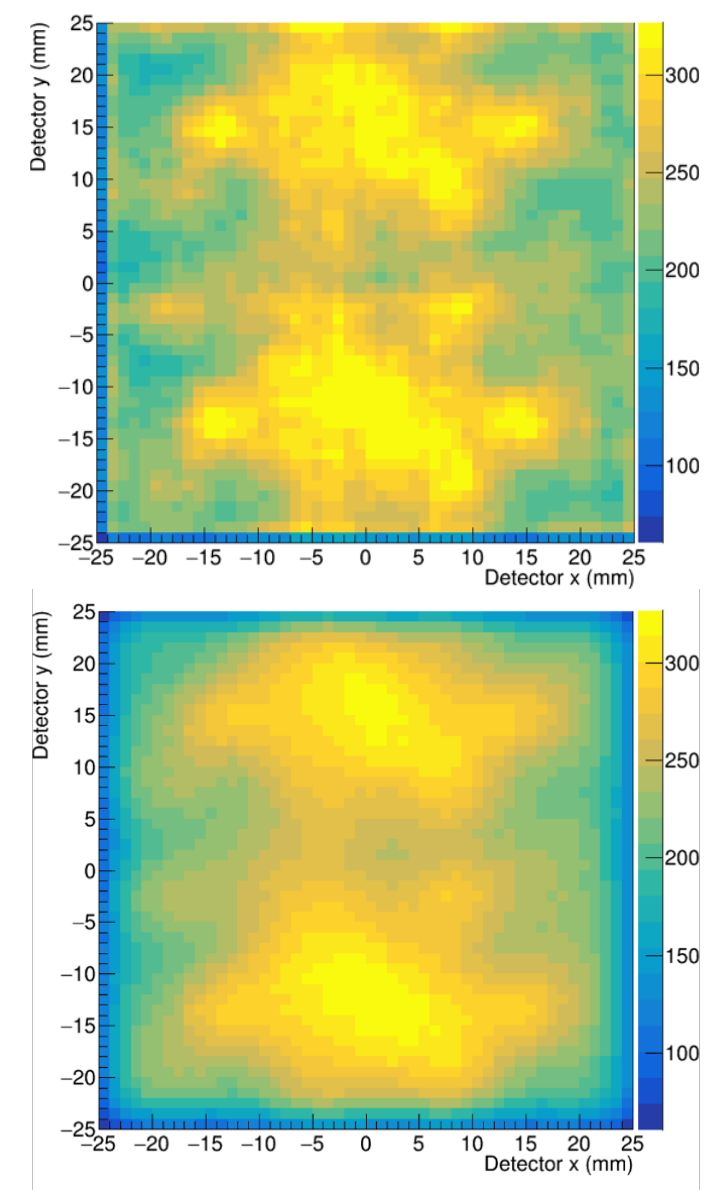}
\caption{Spatial distribution of neutron hits created by a neutron source patter following the "G"
shape. The top pannel shows the ideal response, taken from the simulations. In the bottom one, spatial the response of the CLYC detector has been convoluted with a realistic spatial resolution of 2~mm ($\sigma$).}
\label{fig:PatternsG_CodedMask}
\end{figure}

\begin{figure}[!htb]
\centering
\includegraphics[width=1.05\linewidth]{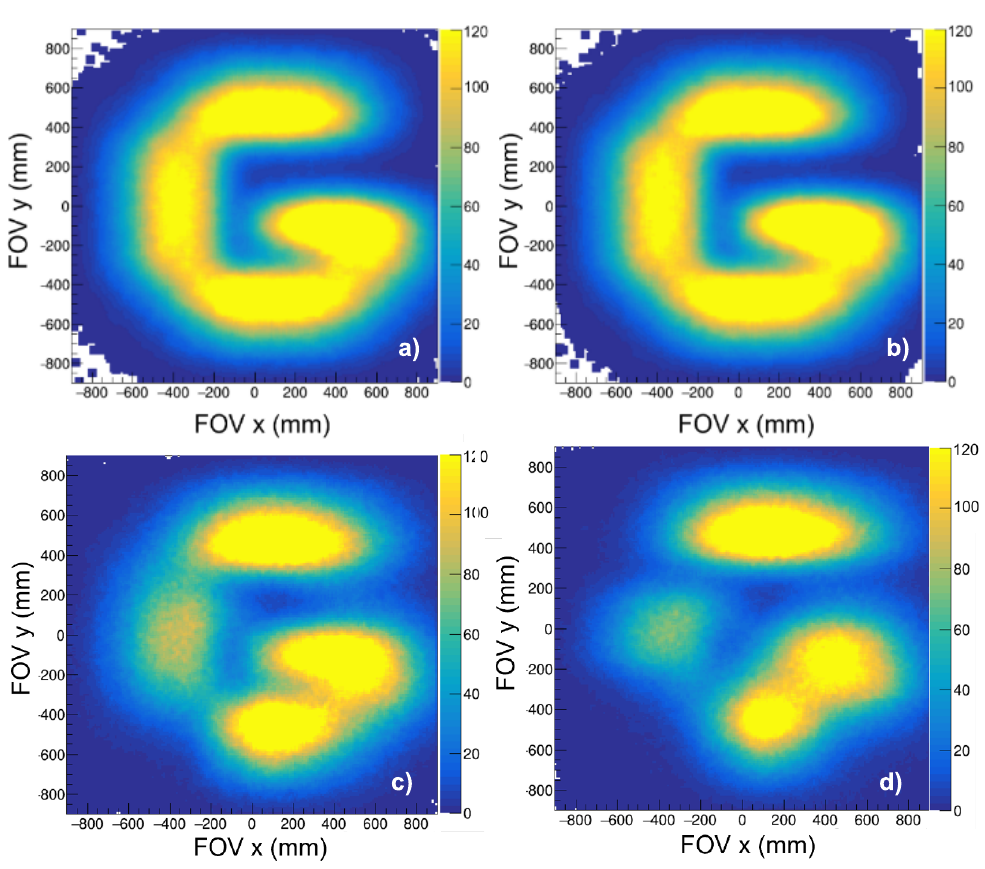}
\caption{Images unfolded for a "G" pattern of thermal neutrons. Top panels: ideal spatial resolution, reconstructed with algorithm from Agostini (a) and  MEM (b). Bottom panels: Spatial resolution of 2~mm ($\sigma$) is added to the response matrix and the image data (c) and only to the image data(d).}
\label{fig:UnfoldedG_CodedMask}
\end{figure}

To investigate the neutron imaging performance employing deconvolution algorithms, the response matrix was built based on the number of counts in each detector pixel (50$\times$50 pixels) in 100 individual simulations of point-like sources placed in a square of 10$\times$10 positions located at a distance of 600 mm from the collimator within the 60$^\circ$ field of view (FOV). The coordinates $x_i$ and $y_j$ spanned from -900 to 900 mm in 200~mm increments. The recorded detector response for each source at these locations formed the foundational database for subsequent reconstruction. Approximately 3$\times$10$^5$ thermal neutron events were recorded for each position.  To evaluate the quality of the reconstructed images, several spatial patterns for the neutron source were created as a linear combination of the 100 individual responses to the simulated point-like source positions. The detector response for each detector pixel $R_j$ for a combined pattern is constructed as follows:
\begin{align}
    \label{eq:DatabaseTensor}
    R_j = \sum_i D_{ij}v_i,
\end{align}
where $v_i$ represents the array of neutron source locations and $D_{ij}$ is the response matrix,  where $i$ denotes the response concerning the source location, and $j$ denotes the detector pixel within the corresponding response. For the initial evaluation of the performance of the unfolding algorithms, we created patterns resembling the shape of the letters of GN-Vision. Figure~\ref{fig:PatternsG_CodedMask} shows the simulated detector response for the spatial pattern "G". The top panel of this figure, which corresponds to the ideal simulated response -- no experimental spatial resolution considered -- leads to the reconstructed images displayed in the top panels of Fig.~\ref{fig:UnfoldedG_CodedMask}. The output of the Agostini algorithm is displayed on the left, while the result obtained with Maximum Entropy is shown on the right. The reconstructed "G" images show a clear contrast and no deformation for both algorithms. To provide the continuous images shown in this figure from the pixelated detector response, a smoothing Gaussian filter was applied.

The next objective was to evaluate the applicability of the coded mask approach with unfolding algorithms to image neutrons in a realistic scenario where the detector response is affected by experimental effects, namely the intrinsic position resolution of the CLYC-SiPM detector. For this purpose, the actual neutron hit coordinates in the CLYC detector, extracted from the \textsc{Geant4} simulations, were convoluted with a Gaussian broadening with a 2~mm ($\sigma$) that matches the actual experimental resolution~\cite{Lerendegui:24_arxiv}. As a result, a blurring of the neutron interaction pattern in the detector is obtained, as shown in the bottom panels of Fig.~\ref{fig:PatternsG_CodedMask}. The impact of the spatial resolution in the reconstructed images was studied considering two different scenarios. The first one aims to model a real case where unfolding is applied to experimental data affected by spatial resolution,  whilst the response matrix is ideal, coming from MC simulations. In the second scenario, a realistic modeling of the intrinsic spatial resolution is added to the response matrix extracted from the MC simulation. The reconstructed images with the Agostini algorithm in these two scenarios are displayed in the bottom panels of Fig.~\ref{fig:UnfoldedG_CodedMask}. The results show the clear degradation of the reconstructed images after experimental effects are taken into account (bottom panels) when compared to the ideal-response images (top panels). When comparing the two experimental scenarios, one observes a clear recovery of the image quality when the modeling of the experimental resolution is added to the response matrix (bottom left) compared to the situation where the response matrix is based only on ideal MC simulations (bottom right).

\begin{figure}[!t]
\centering
\includegraphics[width=0.8\linewidth]{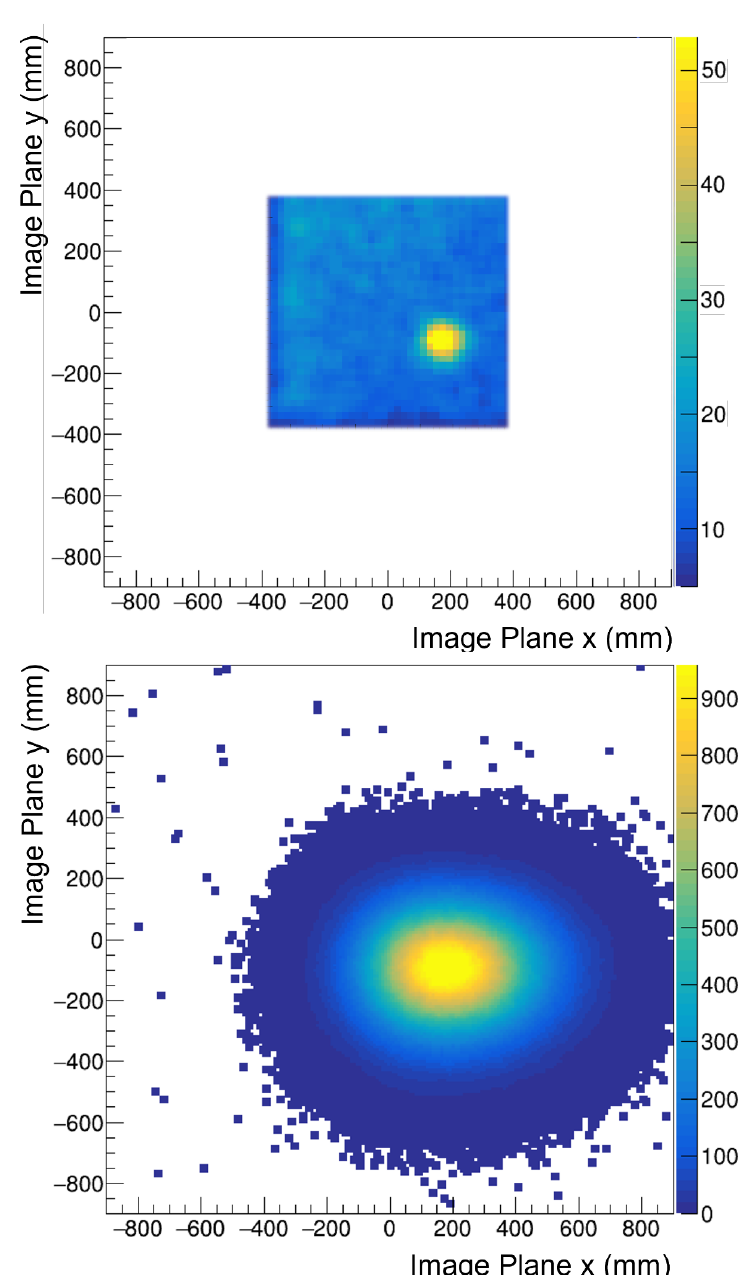}
\caption{Comparison of images of thermal neutron point-like source in the (100, -100, 600) location obtained with the pinhole (top) and the Rank 5 coded mask (bottom).}
\label{fig:Pinhole_vs_CodedMask}
\end{figure}

Upon successful reconstruction of the first thermal neutron images using the coded mask and unfolding algorithms, the impact in efficiency, FOV and resolution was quantified via MC calculations for a point-like neutron source and compared with the simpler pinhole design. As discussed in Section \ref{sec:GNVision}, the pinhole image is essentially an inversion of the hit distribution on the detector, multiplied by a scaling factor. For this comparison, the pinhole focal distance was set to F=40 mm -- like the one used in the aforementioned POC experiments-- , and the aperture radius was D=5 mm, resulting in an open fraction $\rho = \pi/100\approx 0.031$. Last, a pinhole collimator thickness T = 10~mm thick was chosen. A point-like source of thermal neutrons located at (500, -500, 600) was simulated for both the pinhole and the coded aperture mask setups following the aforementioned methodology.

The images reconstructed with the pinhole and coded aperture are shown in Fig.~\ref{fig:Pinhole_vs_CodedMask}. From these images, we have compared the performance of both passive imaging techniques in terms of efficiency, field-of-view and image resolution. The efficiency, calculated from the number of neutron events detected, was found to be 15.6(2) times greater for the MURA rank 5 coded aperture mask than that of the 5~mm diameter pinhole. This difference, related to the difference of open fraction in both collimators,  would be a factor of 4 larger if we compare it to the 2.5~mm diameter pinhole used for the neutron imaging POC experiments discussed in Sec.~\ref{sec:NeutronImaging}. The comparison of the FOV for both collimation systems can be observed in Figure~\ref{fig:Pinhole_vs_CodedMask}, where the images have been scaled so that the FOV is directly comparable. The span of the pinhole FOV -- with the focal distance F=40~mm implemented experimentally in GN-Vision-- is 32$^\circ$, while that of the coded mask would be 60$^\circ$. Last, if we compare the resolution of both images, we find that the performance is significantly worse for the coded mask approach with unfolding algorithms with respect to the simple pinhole with the design parameters D,T,F used in this work. However, the resolution of the coded aperture mask could be improved with a different rank of the coded-aperture mask, more advanced analytical or ML-based algorithms, or a larger set of simulations included in the response matrix.


\section{Conclusions and outlook} \label{sec:Summary}
GN-Vision is an innovative device that is able to simultaneously image $\gamma$-rays and slow neutrons in a single, compact and lightweight device. These features make it of great interest for nuclear safety and control, as well as for nuclear security inspections, where sensitive materials naturally emit both neutrons and \g-rays~\cite{Lerendegui:24}. Moreover, this novel imaging system has potential interest for medical applications, such as proton range monitoring in proton-therapy~\cite{Lerendegui:22_ANPC} or neutron dosimetry in Boron Neutron Capture Therapy~\cite{Lerendegui:24_arxiv}. The proposed device consists of two planes of position-sensitive detectors, based on monolithic LaCl$_{3}$ and CLYC-6 crystals, which exploit the Compton technique for \g-ray imaging. A mechanical lightweight collimator attached to the first plane enables the imaging of slow neutrons ($<$100 eV). 

 In this work, we have given an overview on the development status of GN-Vision, which has focused the experimental efforts on the deployment of its first detection plane, responsible for the neutron imaging capability, based on a CLYC-6 detector with segmented SiPM readout. This work has first summarized the development and characterization of the position-sensitive neutron-gamma discrimination and then reviewed the results of the first experimental validation of the neutron imaging capability of GN-Vision with thermal neutrons using a demonstrator based on a simple $^6$LiPE neutron pinhole collimator attached to the position-sensitive CLYC-6. 

In light of the recent experimental progresses we have explored the prospects for the final GN-Vision design. Accurate MC simulations have been carried out with the aim of evaluating the expected performance of GN-Vision based on the experimental characterization of the individual detectors. The results of the performance study presented herein indicate that the current prototype based on position-sensitive CLYC-6 and LaCl$_3$ detectors would be able to image gamma sources by means of Compton imaging and neutron sources based on the pinhole camera approach, with a remarkable resolution. The predicted neutron image resolution is in good agreement with the one experimentally demonstrated with thermal neutrons in~\cite{Lerendegui:24_AppRadIsot}.

This work has also presented MC-based studies aimed at improving the final design of GN-Vision and overcoming the main limitations of the current prototype, particularly those related to the collimation system. In order to make GN-Vision a more cost-effective solution, we have investigated the impact of modifying the neutron absorbing material in the passive neutron collimator from 95\% enriched $^{6}$LiPE to natural LiPE. The results indicate that the neutron imaging performance, evaluated from the contrast and resolution of the images, will still be acceptable for applications with a fully thermalized neutron spectrum. Another restricting aspect of the first prototype is related to the neutron imaging efficiency, which is limited by the open fraction of the pinhole collimator. Thus, a new neutron-collimation approach based on the coded-mask aperture technique has been investigated, yielding efficiencies that are about one order of magnitude larger than the former pinhole approach. Additionally, the coded-mask approach requires a lesser amount of collimator material, which may have advantages in terms of cost (especially for $^6$LiPE) and Compton imaging performance at low $\gamma$-ray energies.

The next steps in the development of GN-Vision are currently focused on the integration of the dual imaging technique. The Compton imaging of $\gamma$-rays is already at a very high technology readiness level (TRL) following the experience with the former i-TED Compton imager and its adaptations to different applications~\cite{Lerendegui:22,Lerendegui:22_ANPC,Balibrea:22,Lerendegui:24_AppRadIsot,Torres:24,Babiano:24}. This predecessor device was developed on the basis of the PETsys TOFPET2 electronics and, therefore, we aim at keeping the same readout for the LaCl$_3$ crystals of the second (absorber) plane in GN-Vision. The experimental integration of the first detection plane, based on the CLYC-SiPM detector, with the compact PETSys electronics is currently undergoing first experimental tests. However, the CLYC time response differs significantly from that of LaCl$_3$ crystals, which implies that its readout via PETsys electronics is not a straightforward task. Efforts are being made now in order to accomplish this goal, or find alternative solutions that may help to keep it a compact device, while maintaining the satisfactory performance features found thus far, mainly in terms of neutron-gamma discrimination and dual neutron-gamma image resolution and efficiency. 


\section*{Declaration of competing interest}
The authors declare that they have no known competing financial interests or personal relationships that could have appeared to influence the work reported in this paper.

\section*{CRediT authorship contribution statement}
\textbf{J. Lerendegui-Marco:} Conceptualization, Investigation, Methodology, Supervision, Formal analysis, Data curation, Visualization, Writing - original draft, Project administration, Funding acquisition.
\textbf{J. Hallam:} Investigation, Data curation, Visualization, Writing -review \& editing.
\textbf{G. Cisterna:} Investigation, Data curation, Visualization.
\textbf{A. Sanchis-Molt\'o:} Investigation, Data curation, Visualization. 
\textbf{J. Balibrea-Correa:} Investigation, Formal analysis, Software.
\textbf{V.~Babiano:} Investigation, Software. 
\textbf{D.~Calvo:} Investigation, resources.
\textbf{G.~de la Fuente:} Investigation. 
\textbf{B. Gameiro:} Investigation, Software. 
\textbf{I. Ladarescu:} Software. 
\textbf{P. Torres-S\'anchez:} Investigation. 
\textbf{C. Domingo-Pardo:} Conceptualization, Investigation, Supervision, Writing -review \& editing, Project administration, Funding acquisition.

\section*{Acknowledgments}
This work builds upon research conducted under the ERC Consolidator Grant project HYMNS (grant agreement No. 681740) and has been supported by the ERC Proof-of-Concept Grant project GNVISION (Grant Agreement 101113330). We acknowledge funding from the Universitat de Val\`encia through the Valoritza i Transfereix Programme, under grant UV-INV\_PROVAL21-1924580. We also acknowledge funding from the Spanish Ministerio de Ciencia e Innovación under grant PID2022-138297NB-C21. We acknowledge support from the Severo Ochoa Grant CEX2023-001292-S funded by MCIU/AEI. Additionally, the authors thank the support provided by postdoctoral grants FJC2020-044688-I and ICJ220-045122-I, funded by MCIN/AEI/10.13039/501100011033 and the European Union NextGenerationEU/PRTR;  postdoctoral grant CIAPOS/2022/020 funded by the Generalitat Valenciana and the European Social Fund and a PhD grant (PRE2023\_IFIC\_141) and JAE Intro ICU grant from CSIC. Financial support from the Institute Laue Langevin during the experimental campaign is also gratefully acknowledged.





\end{document}